\documentclass[prd,aps,showpacs,floatfix,nofootinbib,twocolumn,preprint,letterpaper,superscriptaddress,reprint]{revtex4-1}
\usepackage{epsfig} 
\usepackage{amssymb} 
\usepackage{float}
\usepackage{amsmath} 
\usepackage{amsfonts} 
\usepackage{color, graphicx} 
\usepackage{bm}% bold math 
 
\usepackage[squaren]{SIunits}
\usepackage{tikz}
\usepackage[symbol]{footmisc}

\begin{document}
\title{Results from phase 1 of the HAYSTAC microwave cavity axion experiment}
\author{L.~Zhong}
\affiliation{Department of Physics, Yale University, New Haven CT, 06511 USA}
\author{S.~Al~Kenany}
\affiliation{Department of Nuclear Engineering, University of California Berkeley, Berkeley CA, 94720 USA}
\author{K.M.~Backes}
\email{kelly.backes@yale.edu}
\affiliation{Department of Physics, Yale University, New Haven CT, 06511 USA}
\author{B.M.~Brubaker}
\affiliation{Department of Physics, Yale University, New Haven CT, 06511 USA}
\author{S.B.~Cahn}
\affiliation{Department of Physics, Yale University, New Haven CT, 06511 USA}
\author{G.~Carosi}
\affiliation{Physical and Life Sciences Directorate, Lawrence Livermore National Laboratory, Livermore CA, 94551 USA}
\author{Y.V.~Gurevich}
\affiliation{Department of Physics, Yale University, New Haven CT, 06511 USA}
\author{W.F.~Kindel}
\affiliation{JILA and the Department of Physics, University of Colorado and National Institute of Standards and Technology, Boulder CO, 80309 USA}
\author{S.K.~Lamoreaux}
\affiliation{Department of Physics, Yale University, New Haven CT, 06511 USA}
\author{K.W.~Lehnert}
\affiliation{JILA and the Department of Physics, University of Colorado and National Institute of Standards and Technology, Boulder CO, 80309 USA}
\author{S.M.~Lewis}
\affiliation{Department of Nuclear Engineering, University of California Berkeley, Berkeley CA, 94720 USA}
\author{M.~Malnou}
\affiliation{JILA and the Department of Physics, University of Colorado and National Institute of Standards and Technology, Boulder CO, 80309 USA}
\author{R.H.~Maruyama}
\affiliation{Department of Physics, Yale University, New Haven CT, 06511 USA}
\author{D.A.~Palken}
\affiliation{JILA and the Department of Physics, University of Colorado and National Institute of Standards and Technology, Boulder CO, 80309 USA}
\author{N.M.~Rapidis}
\affiliation{Department of Nuclear Engineering, University of California Berkeley, Berkeley CA, 94720 USA}
\author{J.R.~Root}
\affiliation{Department of Nuclear Engineering, University of California Berkeley, Berkeley CA, 94720 USA}
\author{M.~Simanovskaia}
\affiliation{Department of Nuclear Engineering, University of California Berkeley, Berkeley CA, 94720 USA}
\author{T.M.~Shokair}
\affiliation{Department of Nuclear Engineering, University of California Berkeley, Berkeley CA, 94720 USA}
\author{D.H.~Speller}
\affiliation{Department of Physics, Yale University, New Haven CT, 06511 USA}
\author{I.~Urdinaran}
\affiliation{Department of Nuclear Engineering, University of California Berkeley, Berkeley CA, 94720 USA}
\author{K.A.~van~Bibber}
\affiliation{Department of Nuclear Engineering, University of California Berkeley, Berkeley CA, 94720 USA}

\date{\today}
\begin{abstract}
We report on the results from a search for dark matter axions with the HAYSTAC experiment using a microwave cavity detector at frequencies between 5.6--5.8$\, \giga\hertz$.  We exclude axion models with two photon coupling $g_{a\gamma\gamma}\,\gtrsim\,2\times10^{-14}\,\giga\electronvolt^{-1}$, a factor of 2.7 above the benchmark KSVZ model over the mass range 23.15\,\textless$\, m_a $\,\textless$\,$24.0$\,\mu\electronvolt$. This doubles the range reported in our previous paper.  We achieve a near-quantum-limited sensitivity by operating at a temperature $T<h\nu/2k_B$ and incorporating a Josephson parametric amplifier (JPA), with improvements in the cooling of the cavity further reducing the experiment's system noise temperature to only twice the Standard Quantum Limit at its operational frequency, an order of magnitude better than any other dark matter microwave cavity experiment to date. This result concludes the first phase of the HAYSTAC program utilizing a conventional copper cavity and a single JPA.
\end{abstract}
\maketitle
\section{Introduction}
\label{sec:intro}
  
The Standard Model of particle physics requires the violation of Charge Parity (CP) symmetry in the strong interaction, which leads to a theoretical neutron electric dipole moment orders of magnitude larger than the current experimental limit. To solve this problem, Peccei and Quinn proposed a solution by which the CP-violating $\theta$ term of the QCD Lagrangian would dynamically relax to its CP-conserving minimum  \cite{b2a, b2b}. Shortly thereafter, Weinberg and Wilczek realized that this mechanism implied the existence of a light pseudoscalar, termed the axion \cite{b3a, b3b}.  Subsequently, it was realized that the properties of the axion and the mechanism by which it would be created in the early universe made it an excellent candidate for the cold dark matter in galactic halos. The axion mass, $m_{ a}$, has historically been taken to be in the range $1\,\mu\electronvolt\lesssim m_{ a}\lesssim1\,\milli\electronvolt$ \cite{b5}. Recent lattice QCD calculations have motivated higher mass axion searches, favoring $m_a\,\gtrsim50\,\mu\electronvolt$ \cite{b6}. Because of its low mass and its very weak interaction with matter and radiation, detecting an axion is very challenging. In 1983, P. Sikivie proposed an experimental axion detection scheme based on the axion-photon conversion \cite{P1983,b7, b8}. The natural conversion rate is very low. For it to be detectable on a reasonable time-scale, this conversion must be resonantly enhanced with a high quality factor microwave cavity in a strong magnetic field. The resulting resonant axion conversion power is:

\begin{figure*}
\includegraphics[width=16cm]{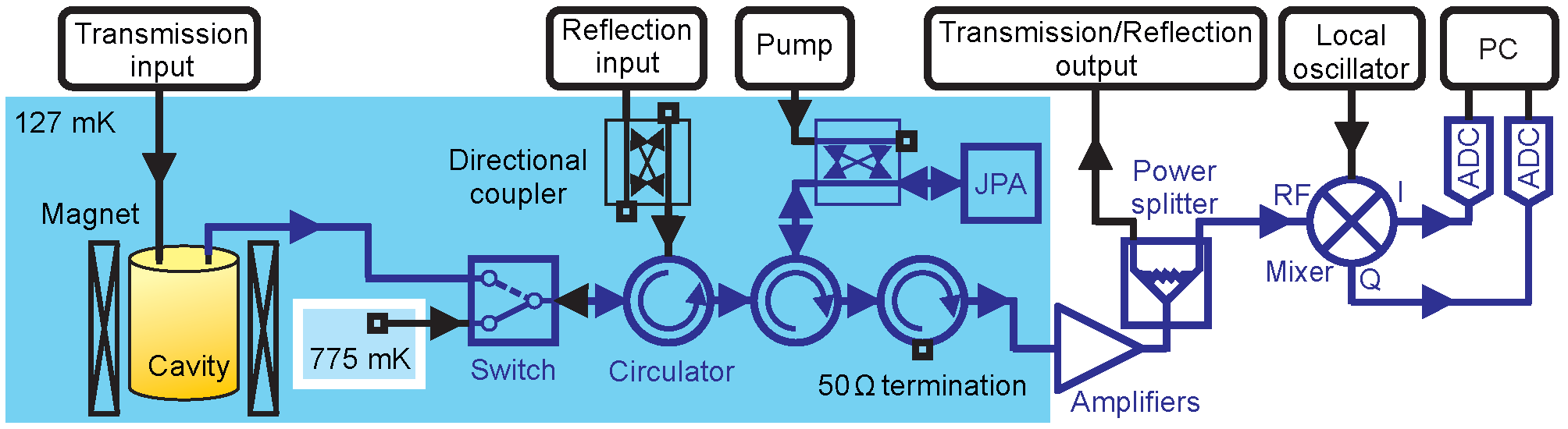}
\caption{\label{circuit}Simplified diagram of the receiver circuit. The detected axion signal would follow the path laid out by the blue arrows. The black arrows indicate the remaining paths. A vector network analyzer (VNA) is used to measure the cavity's frequency response in transmission and reflection.}
%\twocolumngrid 
\end{figure*}

\begin{equation}
P_{ S}=\left( g_{\gamma}^2\frac{\alpha^2\rho_{ a}}{\pi^2\Lambda^4}\right)\left(\omega_{ c}B_0^2V C_{mnl} Q_L \frac{\beta}{1+\beta}\right) \label{first:a}
\end{equation}

Here, $ g_{ \gamma}$ is a model dependent coupling constant, $\alpha$ is the fine structure constant, $\rho_{ a}\,\approx \,$0.45 \giga\electronvolt/cm${^3}$ is the local axion density \cite{b11}, and $\Lambda=\,78\,\mega\electronvolt$ encodes the dependence of the axion mass on hadronic physics.  The physical coupling that appears in the axion-photon Lagrangian is $g_{a\gamma\gamma}\,=\,m_a(\alpha g_{\gamma}/ \pi \Lambda^2)$. The terms in the second set of parentheses in Eq.~1 are experimentally controllable: the coupling coefficient $\beta$\,, unloaded cavity quality factor $Q_0$, loaded cavity quality factor $Q_L\,=\,Q_0/(1+\beta)$, magnetic field $B_0$, cavity frequency $\omega_c$, cavity volume $V$\,, and mode form factor $C_{mnl}$. For $m_{a}\,\approx\, 24\,\mu\electronvolt$, a typical KSVZ model axion with $ g_{ \gamma}~=~-0.97$ \cite{b121, b122}\, gives a conversion power of  $P_{ S}\,\approx\, 5 \times 10^{-24}$ W based on the properties of our detector \cite{b17}. 

The axion has an approximately Maxwellian velocity distribution, and the signal lineshape is given by the corresponding energy distribution. More detailed discussions of the axion signal lineshape can be found in Sec.~VII A in Ref.~\cite{apro}. Because the mass of the axion is unknown, we tune the cavity in discrete steps $\Delta\nu_s \leq\, \Delta\nu_c/2$ and average the cavity noise at each step for an integration time $\tau$. We define the signal-to-noise ratio (SNR) $\Sigma$ as the ratio of signal power to uncertainty in noise power within the signal bandwidth:
\begin{equation}
\Sigma=\frac{P_{ S}}{{k_B} T_{ S}}\sqrt{\frac{\tau}{\Delta\nu_a}}\,.
%SNR=\frac{P_{s}}{P_{n}}=\frac{P_{s}}{{k_B} T_{s}}\sqrt{\frac{\tau}{\Delta\mu}}\,,
\end{equation}
Assuming a phase-insensitive linear receiver, the system noise temperature $T_S$, is given by 
\begin{equation}
\begin{aligned}
\begin{centering}
k_BT_{ S}=h\nu\bigg(\frac{1}{e^{h\nu/k_BT}-1}+\frac{1}{2}+N_{ A} \bigg),
\end{centering}
\end{aligned}
\end{equation}
where the added noise is $N_A\geq \frac{1}{2}$. This method has made axion detection feasible and is the basis of tremendous effort in axion searches. Experiments of this type aim for high magnetic field $B_0$, high $Q_0$, large cavity volume $V$, high form factor $C_{mnl}$, and low system noise temperature $T_{S}$.  

In this paper, we report the results from data runs 1 and 2 of the HAYSTAC (Haloscope At Yale Sensitive To Axion Cold dark matter) experiment. This extends our total coverage to the range 5.6-5.8 GHz with an analysis based on the lab-frame axion lineshape~\cite{PRL}.  Sec.~II describes the experimental apparatus, and Sec.~III describes the improvements on the experiment implemented between data runs 1 and 2. The data analysis and results are described in Sec. IV, with the conclusion in Sec.~V. This completes phase 1 of the HAYSTAC experimental program which utilizes a conventional copper cavity and a single Josephson parametric amplifier (JPA).  The experiment is now being upgraded with a squeezed-vacuum state receiver to improve the sensitivity and scan speed of the search \cite{squeeze}.\\

\section{Experiment}
\label{exp}

The HAYSTAC experiment was first operated in January 2016. Fig. 1 shows a simplified diagram of the receiver circuit. The apparatus is described further in Ref.~\cite{b17}. The experiment employs a 2~liter, high quality factor, tunable microwave cavity maintained at $T_{C} \,=\,127\,$mK. The system is immersed in a strong magnetic field (B\,=\,9\,T) with typical parameters $C_{010}\,\approx\, 0.5$ and $\beta\,\approx\,2$. Galactic halo axions would convert to radio-frequency (RF) photons in the strong magnetic field, and the cavity serves as an impedance matching network that couples the near infinite impedance signal source to a coaxial cable (this can be understood as an extension of the Purcell effect, as originally conceived in Ref.~\cite{purc}). This cable in turn delivers the RF power to a JPA. The experiment requires a narrow-band step-tuned search over frequency for an excess RF noise signal due to axion conversion that would appear as an addition to expected quantum fluctuation noise (along with minimal thermal noise). This tuning (further discussed in Sec.~\ref{tun}) is achieved by the rotation of a copper rod occupying 25\% of the cavity's volume.

To minimize the system noise temperature and allow for \textit{in situ} noise calibration, we designed a receiver that incorporates a near-quantum-limited JPA \cite{b18}, which is a nonlinear LC circuit whose inductance is provided by an array of Superconducting Quantum Interference Devices (SQUIDs), and a microwave switch near the receiver input. The switch can be toggled between a hot load and a cold load. A 50\,$\Omega$ termination thermally anchored to the dilution refrigerator's still plate at $T_{H} = 775\,$mK serves as the hot load, and the cavity serves as the cold load. This toggle setup allows us to incorporate the noise calibrations into the axion search. 

Our preamplifier is composed of the JPA, a directional coupler for the JPA's driving pump tone near its resonant frequency, and a circulator to route the output signal away from the cavity. Two additional circulators isolate the JPA from the cavity and the second stage amplifier, a high electron mobility transistor (HEMT) which is kept at the $4\,$K stage. At room temperature, the signal is further amplified, down-converted to an intermediate frequency (IF) band centered at $780\,\kilo\hertz$, and digitized for analysis. Further details on the experimental setup and signal path can be found in Ref.~\cite{b17}. 

The first data run was carried out over 110\,days and covered the frequency range 5.7--5.8$\,\giga\hertz$. Twenty-three days of rescan focusing on 27 rescan candidates followed this initial run. It was completed in August 2016, with the results and the details of the analysis reported in Refs. \cite{PRL} and \cite{apro}. The run 1 data was analyzed with the axion lineshape in the rest frame, and the exclusion limit in the range of 5.7--5.8$\, \giga\hertz$ was obtained based on this assumption. Rescan candidates from run 1 were re-investigated with a virialized lineshape after run 2. Several technical improvements were implemented between runs 1 and 2. They are described in detail in Sec.~III. Run 2 was carried out over 54 days and finished in July 2017, covering  5.6--5.7$\,\giga\hertz$. Run 2 was followed by 53 days of rescan of potential candidates, where about 75\% of the time was dedicated to candidates from run 1.

\section{Improvements}
The challenges that lead to the technical improvements between the first and second data run are as follows. First, the pulley and Kevlar line system that was used to rotate the tuning rod in the cavity had considerable mechanical hysteresis due to unexpected stiction in the cavity bearings. After each $100\,\kilo\hertz$ tuning step (0.003$^\circ$ rotation), the tuning rod would take 20 minutes to drift slowly to its final position. Second, the tuning rod was supported solely by thin alumina tubes that did not provide a sufficient thermal link to the tuning rod. Because of this, the temperature of the rod remained at 600\,mK, far above the base temperature of 125\,mK.  Finally, the use of thick Cu elements in the construction of the cavity support framework led to major damage of the experiment from the eddy current forces resulting from a superconducting magnet quench during a power outage in March 2016. We now describe our improvements in detail.
\subsection{Piezoelectric Motor Tuning}
\label{tun}

During the run 1, the pulley and Kevlar line system was only used for large frequency steps in order to mitigate the time-dependent drift it caused. Fine stepping was performed by inserting a thin dielectric rod to shift the mode frequency. Unfortunately, the tuning range of the dielectric rod was frequency dependent, and in some regions had no effect at all. Motion of the dielectric rod also generated significantly more heat than the Kevlar pulley system. 

To eliminate the stiction and hysteresis problems, we replaced the pulley and Kevlar line system with an Attocube ANR240 piezoelectric precision rotary motor after the first data run \cite{atto}. The motor is supported by a bracket attached to the experiment frame about 12" above the cavity. The rotary motion is transmitted by 6" long, 0.25" wide brass rods, connected by a corrugated stainless steel flexible shaft coupler. The addition of the new system allowed us to remove the 1.4:1 gear box that coupled the pulley and Kevlar line system to the tuning rod.  

The motor is driven by a sawtooth-voltage electrical signal ($50\,$V amplitude) and draws a high current ($1.5\,$A) when actuated. To ensure low resistance, 28 AWG Cu wires were used to drive the motors from room temperature to the $4\,$K stage, and NbTi superconducting wire was used below the 4~K stage. The motor wires were physically separated from the signal wiring for flux bias current, HEMT amplifier controls, thermometry, heater, and microwave switch to protect the delicate signal wires. The system was tested extensively prior to cool down. To allow for room temperature testing of the motors, the superconducting wires were temporarily replaced with copper wires.

The torque of the Attocube motor is sufficient to move the cavity tuning rod even in the presence of the $9\,$T magnetic field. Empirically the mechanical stiction depends on the direction of rotation. At $9\,$T, the Attocube motor is only able to tune in one direction when the tuning rod is at angles where the stiction is large. This is ok because we carried out the axion search by stepping the cavity frequency in one direction. When necessary, the cavity can be tuned freely in both directions by reducing the magnetic field to $6\,$T. 

The Attocube system generates more heat than the Kevlar pulley system alone, but most of the previous system's heat was generated by the friction cased by the large motions of the dielectric rod. Thus, this upgrade reduced the total heat load of the tuning system.  The Attocube system has provided seamless operation with an acceptable heat load and no observable drift (Fig.~\ref{drift}).  

\begin{figure}
\includegraphics[width=8.5cm]{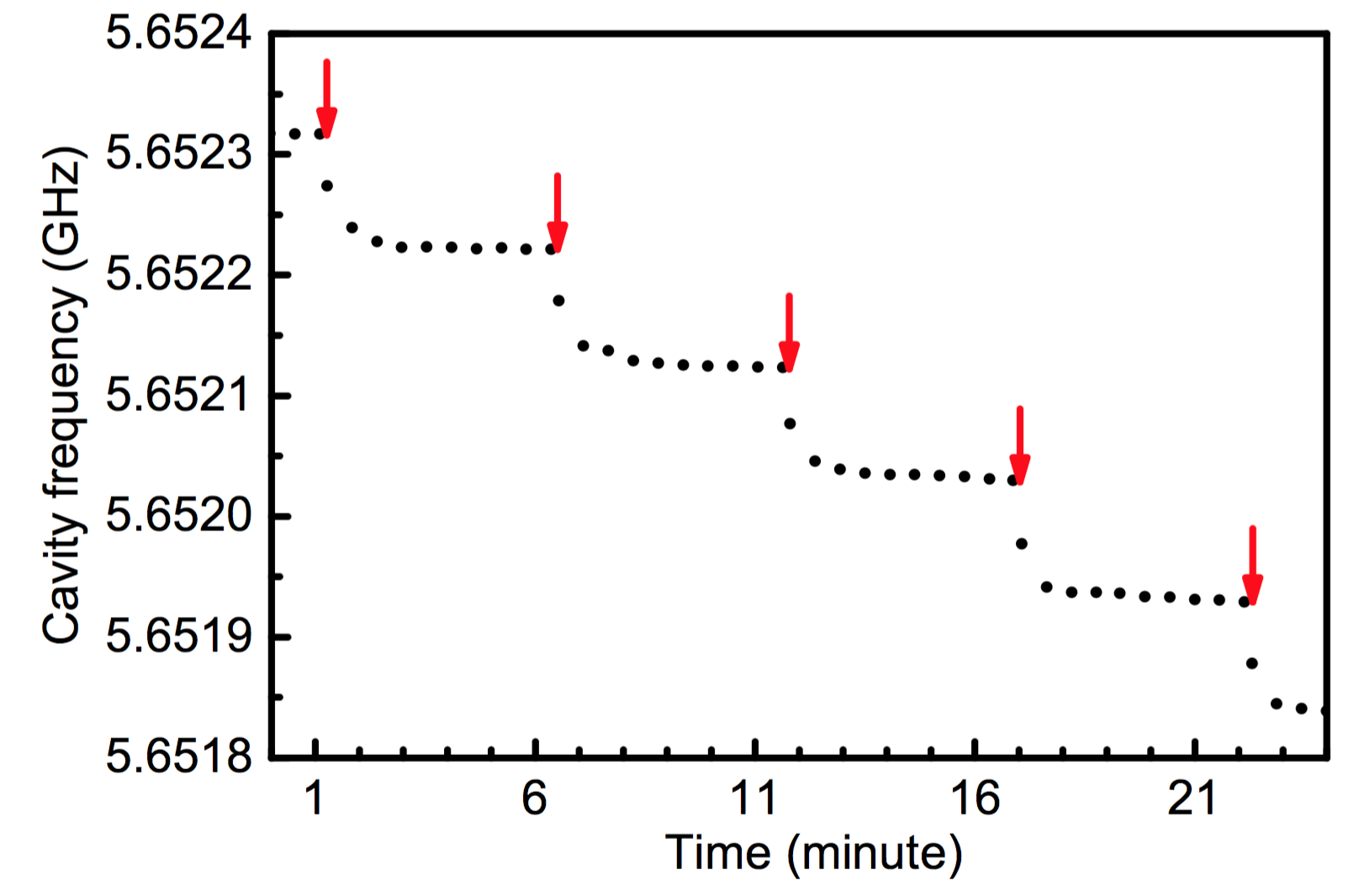}
\caption{\label{drift}Cavity frequency drift between steps of Attocube rotation of the tuning rod with $50\,$V stepping voltage in the presence of a $9\,$T magnetic field. Red arrows indicate stepping. During the data run, there is one minute of wait time between frequency steps.}
\label{drift}
\end{figure}

The addition of the Attocube tuning system has also allowed us to reduce the size of the large notch at $5.704\,\giga\hertz$ in our exclusion limit caused by a cavity mode crossing. We fixed the dielectric rod at two different positions and used the new precision tuning of the Attocube to tune the frequency of the cavity closer to the mode crossing than we were able to before. 

\subsection{Improved Thermal Linkage Between the Tuning Rod and the Cavity}
\label{hotrod}
Prior to this experiment, such a large and uniquely configured cavity had never been coupled to a JPA. During run 1,  the system noise temperature at the cavity resonance was observed to be significantly higher than off resonance. For a thermally well-linked system, we expect the two to be similar. By performing various tests, such as raising the system temperature to the point where the resonance and off-resonance noise levels were nearly equal, it was determined that the excess noise was due to the tuning rod failing to cool to the base temperature of the system. These tests alleviated concerns that the excess noise was due to a spurious interaction between the cavity and the JPA which would have been difficult to eliminate.

The thermal link to the cavity tuning rod comprised two 0.250" outer-diameter, 0.125" inner-diameter polycrystalline alumina tubes, of 4" length each, on either end of the cavity cylinder axis. The only contacts between the tubes and the support frame (which serves as the thermal link to the dilution refrigerator mixing chamber) were bearings that both ensured free rotation of the tuning rod and maintained perfect parallelism between the tuning rod and the cavity body. The contact area between the ball bearings themselves and the inner and outer races of the bearing is vanishingly small by design and did not provide an adequate thermal link. The first attempt to improve the thermal linkage was to glue short brass rods into the external ends of the alumina tubes, using a thermal epoxy, and then connecting those brass rods to the support frame with flexible Cu braids. This solution proved insufficient, as the brass rods were only inserted 0.25" into the tubes, leaving a low-thermal conductivity path of several inches of alumina on either end of the axle.
	
Tests on a nearly identical cavity suggested that 0.125" copper rods could be inserted sufficiently far into the alumina tubes to provide adequate thermal linkage without undue loss of cavity $Q$. Such rods were incorporated into the system and held in place with conductive silver epoxy (Epo-Tek H20E). The rods extend 0.5" beyond the tube where copper braids were soldered onto before gluing. The other ends of the rods serve as connection points to the piezoelectric motor, discussed above.
	
\begin{figure}
\includegraphics[width=8.5cm]{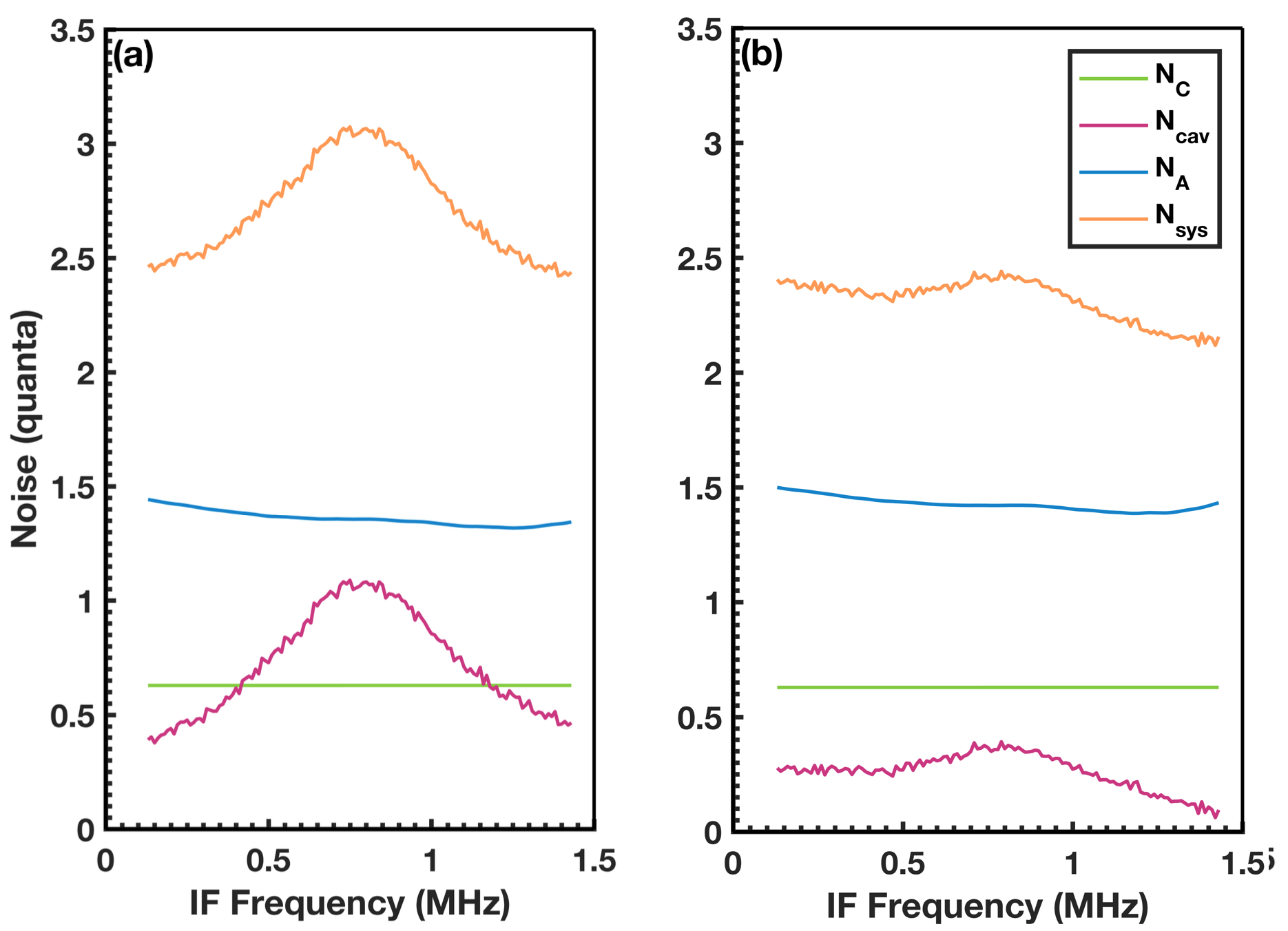}
\caption{\label{noise}Representative noise measurements from (a) run 1 (2016), prior to improving the thermal link with the tuning rod, and (b) the run 2 (2017), after improving the thermal link. $N_C$ (green line) is obtained from thermometry, $N_A$ (blue line) is derived from the average of off-resonance measurements, and $N_{\rm cav}$ (pink line) is the excess noise added by the cavity from a single Y-factor measurement during the data run. $N_{\rm sys}$ (orange line) is the sum of these contributions.}
\label{yfac}
\end{figure}

The copper rods reduce the total system noise photon number at the cavity resonant frequency from around 3 to 2.3 on average, corresponding to a reduction in tuning rod temperature from 600\,mK to 250\,mK (Fig.~\ref{yfac}). Unfortunately,  the cavity $Q$ was reduced by about 40\%. We now believe that the failure to achieve the original $Q$ was due to the construction of the tuning rod's axle for the cavity used in actual running.  It prevented the copper rods from being placed in the optimum position.  This will be corrected for future runs, where we predict no additional thermal contribution from the rod, and essentially no diminution of $Q$. 

The improved thermal link reduced the time the system takes to cool from the first condensing of the mixture to base temperature from over six hours to under one hour. Without the copper tubes inserted, the alumina tubes' weak thermal link was a bottleneck in the cooling process and maintained a substantial heat load. When the tuning rod reached 600~mK, the alumina tubes became effective thermal insulators, keeping it from reaching 125~mK. After this quasi-equilibrium was established, we saw no discernible decrease in thermal noise level over months of operation, implying a time of perhaps years for the tuning rod to significantly cool beyond this point. We have yet to identify the remaining source of excess thermal noise (250~mK compared to a system temperature of 125~mK) which is likely a further issue with the thermal link of the tuning rod. 

\subsection{ Copper Plated Stainless Steel Thermal Links and Shields}
The original design incorporated several massive OFHC copper components. A magnet quench during a university-wide power outage caused significant damage to the experiment. The cavity is made of copper plated stainless steel and has a high thermal conductivity. Readings from the thermometers on the cavity top and bottom indicate that there is less than 30 seconds time delay between a change in temperature at the top, and a subsequent change at the bottom. The rapid change of the cavity temperature at its bottom (far end from thermal link) led us to conclude that heavily plated stainless steel could be used to construct effective thermal radiation shields while minimizing the amount of copper in the system.

The damaged still-temperature thermal shield was replaced with a stainless steel shield plated with 0.002" copper. This new shield has been sufficient, with no obvious excess heat load at the mixing chamber level. 

\begin{figure*}
\includegraphics[width=\textwidth]{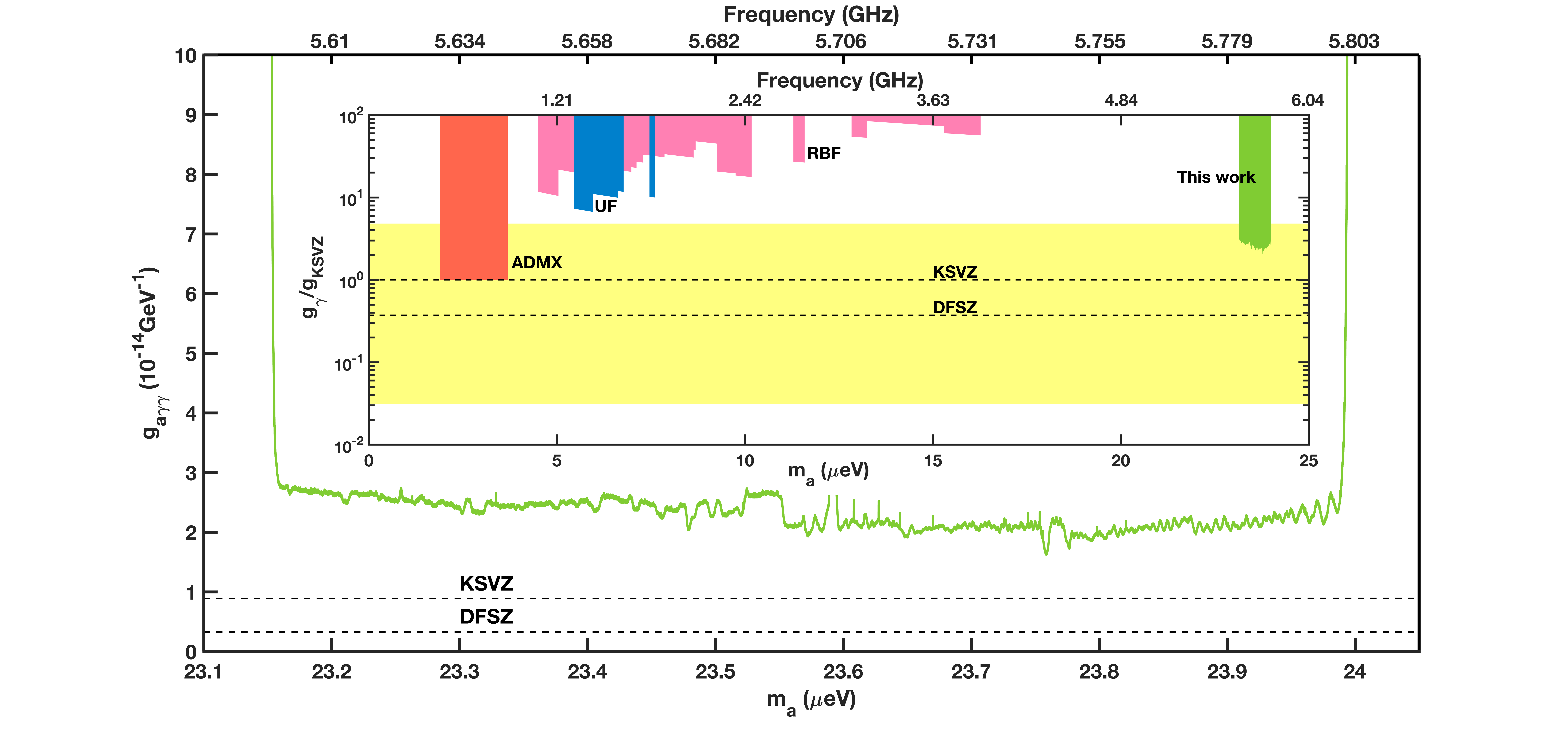}
\caption{\label{fig5}Our exclusion limit at 90\% confidence. Green represents this work combined with our previous results presented in Ref.~\cite{PRL}. Red represents previous cavity limits from ADMX \cite{b91, b92, b93}, pink represents results from Brookhaven \cite{b20}, and blue represents results from the University of Florida \cite{b21}. The axion model band is shown in yellow \cite{b14}. The KSVZ \cite{b121, b122} and DSVZ \cite{b131, b132} couplings are plotted as dashed lines.}
\label{totexc}
\end{figure*}

\section{Data Analysis and Results}
The combined data from run 1 and run 2 covers the frequency range 5.6--5.8$\, \giga\hertz$. The data runs resulted in a total of 10406 raw subspectra, of which 10090 were used for the analysis presented here. The remaining 316 were rejected due to their poor JPA gain stability, cavity frequency drift, proximity to cavity mode crossing, etc. Each sub-spectrum covers a 1.3 MHz analysis band with resolution of $\Delta \nu_b\,=\,100\,\hertz$.   Here we give a brief description of the analysis. The analysis procedure is detailed in Ref.~\cite{apro}. 
%look at this par a lot 
The final limit shown in Fig.~4 is obtained by combining the 10090 selected subspectra in a weighted sum that maximizes the SNR. The subspectra are aligned by their IF frequency and averaged to extract the average shape of the spectral baseline. Aligning them in this manner allows us to cut IF bins that have been compromised by narrow IF spikes from the analysis. Next, the average shape of the spectral baseline is removed from each raw subspectrum. The remaining baseline structure is removed by dividing out a Savitzky-Golay (SG) fit and subtracting 1. In the absence of an axion, each raw subspectrum is now a dimensionless processed subspectrum described by the same Gaussian distribution. This Gaussian distribution has a mean of $\mu$ = 0 and standard deviation of $\sigma = 1/\sqrt{\tau\Delta\nu_b}$. To put the the raw subspectra in units of watts, each raw subspectrum is multiplied by the average noise power per bin. This also undoes the suppression of any potential signal that would appear in a particularly noisy bin. Now to scale the raw subspectra such that an axion present in any bin would have the same value, we divide by the Lorentzian axion conversion power profile. The expected axion power is different across the RF frequency spectrum and depends on the cavity quality factor $Q$, coupling factor $\beta$, mode form factor $C$, and the cavity transmission. In order to form the combined spectrum, corresponding RF bins in different spectra are added together with maximum likelihood (ML) weighting. Groups of ten neighboring bins are then added together with an extension of the ML method reducing the resolution of the spectrum from $\Delta\nu_b\,=\,100\,\hertz $ to $\Delta\nu_b\,=\,1\,\kilo\hertz$. Next, overlapping groups of nine neighboring bins are added together, this time taking into account the expected axion lineshape. In each step of this process, the standard deviation of each sample is also scaled accordingly. 

We set a threshold in the combined spectrum based on a predetermined confidence level and target axion coupling. This allows us to select frequencies passing this threshold as rescan candidates. We set a frequency-independent target SNR $\Sigma$=4.78~$\sigma$, implying a frequency-dependent target axion coupling. If an axion with the target coupling exists at a certain frequency, it will appear in the grand spectrum with a mean of 4.78~$\sigma$ and a standard deviation of $\sigma$. Based on the target SNR, we can set a power threshold to determine rescan candidates. This threshold must be lower than $\Sigma$ to ensure that a real axion signal would be flagged as a rescan candidate with high confidence. If the threshold were to be set equal to $\Sigma$, then an axion with SNR $\Sigma$ would have a power excess that is Gaussian-distributed around $\Sigma$, and only have a 50\% confidence limit (C.L.) for appearing above the threshold. 27 rescan candidates, defined as frequency ranges with normalized power exceeding 3.5~$\sigma$ (90\% C.L.) were identified. This is consistent with the number of rescan candidates expected by simulating Gaussian white noise subjected to the same co-adding procedure.

To determine whether each rescan candidate is due to a statistical fluctuation or a persistent RF noise excess, such as an axion to photon conversion, we must set a target rescan SNR $\Sigma_{r}$ for the rescans associated with each data run. As discussed in Sec.~\ref{hotrod}, the improvements in the thermal coupling of the alumina rod reduced the system noise temperature by 20\% but decreased the $Q$ by 40\%. This $Q$ degradation implies that a different $\Sigma_{r}$ is appropriate for the two data sets. The rescan integration time $\tau$ required to achieve $\Sigma_{r}$ is frequency dependent. It is also depends on $\Sigma_r$ as follows:
\begin{equation}
\label{times}
\tau\propto\bigg(\frac{T\Sigma_r}{Q_L}\bigg)^2
\end{equation}

From Eq.~\ref{times}, we can see that the decreased $Q$ leads to a significantly increased integration time required to reach the same SNR as in the initial scan. The longer the integration time at a certain frequency, the more pronounced the baseline systematics become in the shape of the data. This limits the amount of time we can take data at each frequency. Accounting for this effect, we chose values of $\Sigma_r$ = 4.53~$\sigma$ and $\Sigma_r$ = 5.1~$\sigma$ for the first and second scans, respectively. From $\Sigma_r$, cavity parameters, and parameters measured in the initial scan, the rescan integration time per frequency can be determined.  The data collected is then analyzed in a similar method as detailed above with different filtering parameters. This process is further detailed in Sec. IX (Rescan Data and Analysis) in Ref.~\cite{apro}. Of the 27 rescan candidates, four passed this second threshold and were at the frequencies 5.72648~GHz, 5.71761~GHz, 5.71652~GHz, and 5.66417~GHz. These four rescan measurements were repeated, and they did not pass the threshold a third time. We see this by adding the data from the rescan and the second rescan for these candidates. This raises the effective integration time which increases the threshold. The excess power at these frequencies is now clearly below the new higher threshold. 

Of the 27 rescan candidates, the spectra around 5.79697~GHz, 5.76952~GHz, 5.76318~GHz, 5.75986~GHz, 5.74421~GHz, 5.74418~GHz, and 5.73688~GHz (all from the frequency range covered by the first data run) exhibited non-Gaussian statistics. During the initial run, these frequencies had Gaussian spectra. Because the non-Gaussian behavior only appeared after the thermal coupling problem was fixed, it is believed that the added Cu rod acts as an antenna and couples the spurious RF signals into the cavity. This problem is more prominent during rescans than during data runs because the cavity sits at one frequency taking data for a longer time duration. During the second data run, there were also six narrow features ($\leq$ axion width). It is believed that these signals were also coupled into the cavity through the Cu rod. These were proven not to be from an axion signal by taking measurements without the magnetic field or by taking measurements off cavity resonance. 

We report the limit for $g_{\gamma}$ based on the combined axion search data from runs 1 and 2 in Fig.~\ref{totexc}. We do not have a definitive candidate at this time. 
\section{Conclusion}

We report results from the first haloscope axion detector to achieve sensitivity to cosmologically relevant couplings at masses above 20\,$ \mu$eV. The difficulty of reaching higher axion masses comes from the fact that the effective volume $VC_{mnl}$ of the cavity in which axion coupling can occur falls off rapidly with increasing frequency. Despite the difficulty of working in this mass range, we were able to reduce the total noise to 2.3 times the standard quantum limit, and set an exclusion limit of $|g_\gamma|\gtrsim 2.7\times |g_\gamma^{\mathrm{KSVZ}}| $ over the range 23.15\,\textless$\, m_a $\,\textless$\,$24.0$\,\mu\electronvolt$. This sensitivity is already well into the space of plausible model couplings and best-estimate halo densities.  That such a small experiment, of order 1\% of the volume of prior experiments \cite{b20} is discovery-capable, is a remarkable achievement, and is primarily due to dramatic advances in amplifiers enabling operation very near the quantum limit. The experiment was further refined by the solution to the thermal coupling problem, addition of the Attocube tuning system, and improved shielding. This concludes the first phase of HAYSTAC's operation. The next phase will include upgrades to the analysis and the implementation of a squeezed stated receiver which will allow us to push down even further in sensitivity \cite{squeeze}. 

\section{Acknowledgements}

This work was supported by the National Science Foundation, under grants PHY-1362305 and PHY-1607417, by the Heising-Simons Foundation under grants 2014-181, 2014-182, and 2014-183, and by the U.S. Department of Energy through Lawrence Livermore National Laboratory under Contract DE-AC52-07NA27344. We gratefully acknowledge the critical contributions by Matthias B{\"u}hler of Low Temperature Solutions UG to the design of and upgrades to the cryogenic system. 

\newpage

\bibliography{PRD1.bib}

\end{document}